\documentclass[prb, twocolumn]{revtex4-1}
\usepackage{graphicx}
\usepackage{latexsym}
\usepackage{amsmath}
\usepackage{amssymb}
\usepackage{epstopdf}
\usepackage{enumerate}
\usepackage{setspace}   
\usepackage{dcolumn}
\usepackage{bm}
\usepackage{setspace}   
\usepackage{slashed}

\begin{document}
\title{   
Skyrmions with quadratic band touching fermions: \\
A way to achieve charge $4e$ superconductivity.
}

\author{Eun-Gook  Moon}
\affiliation{Department of Physics, University of California,
Santa Barbara, CA 93106}

\date{\today}

\begin{abstract}
We study Skyrmion quantum numbers, charge and statistics, in $(2+1)$ dimension induced by quadratic band toucing(QBT) fermions.
It is shown that induced charge of Skyrmions is twice bigger than corresponding Dirac particles' and their statistics are always bosonic. 
Applying to the Bernal stacking bi-layer graphene, we show that Skyrmions of quantum spin Hall(QSH)  are charge $4e$ bosons, so  their condensation realizes charge $4e$ superconductivity(SC). 
The phase transition could be  a second order, and one candidate theory of the transition is $O(5)$ non linear sigma model(NLSM) with non-zero Wess-Zumino-Witten(WZW) term. 
We calculate renormalization group beta function of the model perturbatively and propose a possible phase diagram. 
We also discuss how QBT fermions are different from two copies of Dirac particles.  
\end{abstract}

\maketitle


{\it Introduction :} 
Electronic strucutres with helicity such as ones of Dirac particles and quadratic band touching(QBT) fermions in two-spatial dimension receive a lot of attention especially with the discovery of graphene. \cite{novoselov}
The helicity and its associated Berry phase play an important role in quantum mechancal phenomena such as half-integer Quantum Hall effects in graphene.\cite{neto,kim}
In particular, QBT fermions lead more interesting and puzzling physics due to a stronger interaction effect and are not well understood yet. \cite{lau,kim2,yacoby}
Not only graphene but also some spin liquids and topological insulator also have QBT fermions structures. \cite{fu,sun,xu,moon}

In (2+1) dimension, there is another common and interesting excitation, Skyrmion of $O(3)$ vector order parameters. \cite{shin,han}
For example, the Neel order phase has Skyrmions because its ground state manifold is a two sphere due to its broken symmetry structure, $SU(2)/U(1)=S^2$. 
Basically, a Skyrmion number is a wrapping number of a given configuration on a two dimensional space with identified boundaries. 

One natural question with the two excitations, QBT fermion and Skyrmion, is how they interact with each other.
In this paper we answer the question focusing on Skyrmions' induced quantum numbers by QBT fermions. Extensive research already exists on Skymions with Dirac particles. \cite{abanov,abanov_SC}

Our goal  is two-fold. 
First, we show Skyrmions receive quantum numbers by interacting with QBT fermions, which are doubled compared to Dirac particles'.
In addition, we apply the mechanism of the induced quantum numbers of Skyrmions to the Bernal stacking bi-layer graphene structure and 
study an exotic quantum phase transition in the bi-layer graphene.  
We report two findings.  First, quantum spin hall phase in the bi-layer graphene is connected to charge $4e$ SC  by QSH Skyrmions condensation.  
And second, the phase transiiton could be a second order and we propose one candidate theory, $O(5)$ non linear sigma model with the Wess-Zumino-Witten term.

In terms of charge $4e$ SC, one can see that it is almost impossible to realize it with the conventional BCS theory 
because two pairs of charge $2e$ order parameters have a better chance of forming than charge $4e$ SC.
Therefore, one key factor to achieve charge $4e$ SC is how to prohibit charge $2e$ order parameters. 
One possibility of charge $4e$ superconductivity was suggested  by Berg. {\it et, al.}. \cite{erez}
Main idea of their study is to prohibit $2e$ pairings thermally to make $4e$ superconductors,  so they only find non-zero temperature charge $4e$ SC. 
Our research is a significant departure from the previous one by Berg 
in that our $4e$ SC is a ground state at zero temperature.

Abanov and Wiegmann studied a non-BCS way to achieve superconductivity focusing on acquired quantum numbers of topological objects interacting with Dirac fermions.\cite{abanov_SC}
If a quantum number-endowed topological object is a boson, it is possible to condense the object, and it induces quantum phase transition as a textbook example in the two dimensional $XY$ model. 
Applying the idea to monolayer graphene, Grover and Senthil argued that quantum spin hall phase in monolayer graphene can be connected 
to charge $2e$ superconductor through the $CP(1)$ deconfined quantum criticality.\cite{grover}
Below we show how QBT fermions change Skyrmions' properties and associted Quantum phase transition.

{\it Model Hamiltonian : } 
Before going into QBT fermoins, let us briefly review results of Dirac particles. \cite{abanov,abanov_SC,grover}
Hamiltonian consists of Dirac fermions with $O(3)$ vector order parameter.
\begin{eqnarray}
H_{Dirac}= {\psi}^{\dagger} \left[k_x \sigma^x  + k_y \sigma^y + g \sigma^z \vec{n} \cdot \vec{\tau}\right] \psi.
\label{Hamiltonian}
\end{eqnarray}
, where $\sigma^i, \tau^j$ are different kinds of Pauli matrix, so it is necessary to introduce at least four component spinors.
It is well known that the $O(3)$ vector fields allow non-trivial topological configuration, Skyrmion, due to $\pi_{2}(S^2) =Z$. 
Its topological current is $J_{\mu} =\frac{1}{8 \pi} \varepsilon_{\mu \nu \lambda} \vec{n} \cdot (\partial_{\nu}\vec{n} \times \partial_{\lambda} \vec{n} )$
and charge is spatial integration of its time component, $Q^{Skyr}=  \int d^2 x \frac{1}{4 \pi} \vec{n} \cdot (\partial_{x}\vec{n} \times \partial_{y} \vec{n} )$. 

It was shown that the induced fermionic current, $<j^{ferm}_{\mu}>$, is proportional to  Skyrmion  current and its charge is {\it exactly} determined by Skyrmions number upto number of flavors. 
\begin{eqnarray}
&&Q^{ferm} = N_f Q^{Skyr} \label{dirac_c}
\end{eqnarray}
Also, it was well known that Skyrmions' statitical properties could be read off by examining  two Skyrmions exchange property or measureing a phase of $2\pi$ rotation.
Those information is contained in the so-called Hopf term, which corresponds to $\pi_{3}(S^2) =Z$ mathematically. 
\begin{eqnarray}
&&   S_{Dirac}^{Hofp} = i N_f \pi \cdot {\rm sgn}(g)  \int_x A_{\mu} J^{\mu} \label{dirac_h} \nonumber
\end{eqnarray}
, where $N_f$ is the number of flavors and the gauge field($A_{\mu}$) is defined by the relation, $J_{\mu} = \varepsilon_{\mu \nu \lambda} \partial_{\nu} A_{\lambda}$.

Massage of the above discussion is manifest. Skyrmions of order paprameters receives charge and statistics by Dirac fermions.
Dirac particles with even number flavors make Skyrmion bosonic charged particles and odd number's make it fermionic charged paricles. 
Notice that there is one dimensionful coupling constant($g$) but  it does not appear in our final results except its sign. 
It is a signal of topological origin of induced quantum numbers of Skyrmions. 

Now, let us consider quadratic band touching fermions interacting with $O(3)$ vector.
In general, Hamiltonian is
\begin{eqnarray}
H= \psi^{\dagger}\left[ \frac{ (k_x^2 -k_y^2) \sigma^x  +(2 k_x k_y ) \sigma^y}{2m}+ g \sigma^z \vec{n} \cdot \vec{\tau}\right] \psi . \nonumber
\end{eqnarray} 
This is a generalization of the Eqn. \ref{Hamiltonian} from the $p$ wave helicity to the  $d$ wave helicity with introducing one more dimensionful scale, effective mass($m$).
Obviously, the Lorentz symmetry is broken by quadratic dispersion relation. 
Such systems can be realized, for example, in bi-layer graphene or in the hole-doped semiconductors with the inversion symmetry. 

One way to see induced charge of Skyrmion is to evaluate fermion's energy spectrums under Skyrimon background. 
Turning on Skyrmion configuration adiabatically, one can show there are zero energy level crossing states. 
For one spatial dimension with $O(2)$ symmetry, Yao and Lee showed there were two level crossing for the QBT. \cite{yao}
In principle, one can do similar calculation in two spatial dimension.
Instead, we use a field theoritic method, the graident expansion, to see both Skyrmion's charge and statistics also.
Similar calculation method has been applied to Dirac particles in various literatures. \cite{abanov}

\begin{figure}
\includegraphics[width=2.0 in]{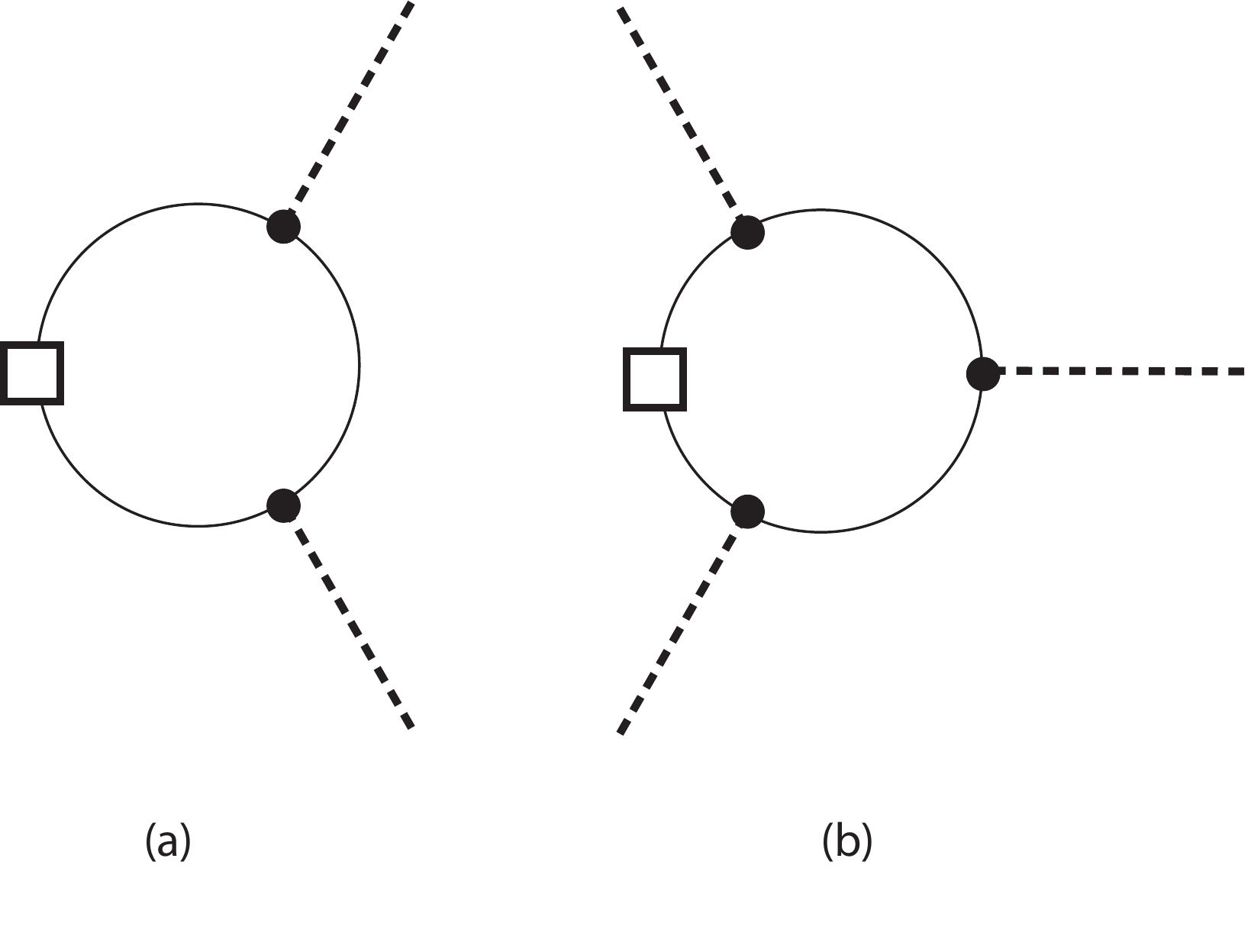}
\caption{ Induced current(a) and Hopf term(b) diagram. The plain line is for the propagator.($G_0$) The square vertex is for the variation of the $D$ operator.($\delta D D^{\dagger}$) The circle vertex is the coupling between the fermion and the order parameter variations.($M$) }
\label{Loop}
\end{figure}

For QBT fermions, broken Lorentz symmetry makes the expansion little more tedious but it is straightforward to do if one only keeps the lowest term. 
Partition function is 
\begin{eqnarray}
\mathcal{Z} = {\rm Tr} (e^{-\int d^2 x d \tau \mathcal{L}}) = e^{-S_{eff}} \quad , \quad  \nonumber 
\end{eqnarray}
, where the Lagrangian is 
\begin{eqnarray}
\mathcal{L} &\equiv& \psi^{\dagger}D\psi =  \psi^{\dagger}  \partial_{\tau} \psi +H . 
\end{eqnarray}
One key step of the gradient expansion is to consider variation of the effective action.
\begin{eqnarray}
&&\delta S_{eff} = -{\rm tr} (D^{-1} \delta D) =  -{\rm tr} (\delta D D^{\dagger} (D D^{\dagger})^{-1} ) .
\end{eqnarray}  
Following commutation relations are useful. 
\begin{eqnarray}
&&[k_x^2-k_y^2, \hat{n}] = 2 (-i) (k_x \partial_x \hat{n} -k_y \partial_y \hat{n}  ) + (\partial_x^2-\partial_y^2) \hat{n} \nonumber \\
&&[k_x k_y + k_y k_x , \hat{n} ]= 2 (-i)( k_x \partial_y \hat{n} + k_y \partial_x \hat{n}) +2 \partial_x \partial_y \hat{n}
\nonumber
\end{eqnarray} 
If we keep only the lowest order terms, we get
\begin{eqnarray}
&&DD^{\dagger} \equiv G_0^{-1} +M =\omega^2+(\frac{k^2}{2m})^2 +g^2 + M \nonumber \\
&&M = g \left[ \sigma^z \partial_{\tau} \hat{n} + \frac{k_y \sigma^x- k_x\sigma^y}{m}  \partial_x \hat{n} +  \frac{k_x \sigma^x+ k_y\sigma^y}{m}  \partial_y \hat{n}  \right] \nonumber 
\end{eqnarray}
With these expansion, one can do perturbative calculation with Green's function, $G_0$ with perturbation, $M$.
In Fig. \ref{Loop}, we illustrate two diagrams for induced current(a) and Hopf term(b) calculations.
Note that to calculate induced currents it is necessary to couple external gauge field as in an usual way, ($k_{\mu} \rightarrow k_{\mu}- a_{\mu} $).
By evaluating the first diagram, we find  
\begin{eqnarray}
Q^{ferm}= 2 N_f Q^{Skr}.
\end{eqnarray}
Here, we put $N_f$ back to the formular for notational purpose.

Few remarks are in order. 
First, there is a factor two compared with the Dirac fermion's of Eqn. \ref{dirac_c}.
The induced charge of the Skyrmion by QBT fermions is doubled compared to the Dirac case.
Second, the result indicates that Lorentz symmetry is not a crucial factor for topologically induced currents. 
In other words, the vacuum of the non relativistic fermion with the topological background could be  polarized and its polarization is twice bigger. 
Also, one notice that there is no effective mass dependence in the current calculations.
This fact is another signature of its topological origin.
Otherwise, one expect that the effective mass term would appear in momentum integration.

To complete induced quantum numbers, let us study statistics of topological defects. 
It was shown by Wilczek and Zee that the Hopf term of the non-linear sigma model, so-called the ``theta'' term, describes the statistics of the Skyrmion.\cite{wilczek}
\begin{eqnarray}
S^{Hopf} = i \theta \int_{x} A_{\mu} J^{\mu}
\end{eqnarray}
, where $\theta$ term is determined by microscopic models. 
If we consider the rotation(or exchange) of the soliton, then its current naturally induces the non-trivial imaginary factor depending on its spin(J).
\begin{eqnarray}
S \rightarrow S+ i \theta \quad ( S \rightarrow S+ i 2 \pi J)
\end{eqnarray}
So the Skyrmion spin is $J= \frac{\theta}{2 \pi}$, and $\pi$($2\pi$) corresponds to fermion(boson).
Other values indicate anyonic statistics of the Skyrmion. 

Following the same steps as in the current calculaiton, it is straighforward to obatin $ \delta S^{Hopf}_{quad} $ using the second diagram of Fig. \ref{Loop}(b).
We find that  imaginary part of effective action's variation corresponding to the Hopf term is  
\begin{eqnarray}
 \delta S^{Hopf}_{quad}&=&  2i \frac{{\rm sgn}(g) }{32 \pi}   \int_x   \varepsilon^{\mu \nu \lambda} {\rm tr} [\delta \hat{n} \hat{n} \partial_{\mu} \hat{n}  \partial_{\nu} \hat{n}  \partial_{\lambda} \hat{n}]
\end{eqnarray}
, where  $\hat{n} \equiv \vec{\tau} \cdot \vec{n}$ is used.

It is obvious that the variation becomes zero, which naively looks no Hopf-term contribution. 
However, it is well known that the Hopf term caluculaiton is more subtle due to its non-trivail homotopy group, $\pi_3(S^2) = Z$.
In other words, the zero variation is an artifact of our perturbative method.
To overcome this difficulty, we use the embedding method. \cite{witten, abanov}
Main idea is to extend  the order parameter group, $CP^{1} \sim S^2 $, to enlarged $CP^{M}$ to avoid the ambiguity.
\begin{eqnarray}
\hat{n} \equiv \vec{n} \cdot \vec{\tau} \rightarrow 2 z z^{\dagger} -1   \label{na}
\end{eqnarray}
with $z^{t} = (z_1, \cdots , z_{M+1})$ and $z^{\dagger} z =1$. 
Due to the fact, $\pi_3 (CP^M) =0$, for $M>1$ it is safe to use a perturbation theory. 
Variation of the effective action has no ambiguity and our perturbative calculation is well-defined. 
Then, we take the limit $M \rightarrow 1$ and read off the induced Hopf term.
And the induced Hopf term is 
\begin{eqnarray}
 S^{Hopf}_{quad}
 &=& 2  N_f \pi i\,\, {\rm sgn}(g) \int_{x} A_{\mu} J^{\mu}.
\end{eqnarray}
We put $N_f$ back to the final forms for notational purposes. 
As we can see, the factor two again appears in the Hopf term calculation, which is consistent with the  charge current calculation. 
Therefore, Skyrmions with the QBT fermions are always bosons and their charge is determined by number of flavors($2N_f$). 
Again, the dimensionful effective mass does not appear due to its topological origin.

How do we understand Skyrmion charge doubling effect in QBT fermions?
One way to understand is to add small perturabation which do not break other symmetries such as gauge potential disorder with strengh(q). 
Then, it is easy to see the QBT point at the minium splits into two Dirac points below the energy sclae corresponding $(q)$. 
If only low energy properties are dominant to determine topological charges, we can just add two Dirac particles to make QBT fermions, which explains the doubling effect. 
It seems the argument that QBT is equivalent to two Dirac particles is compelling, but they show different fermionic mass structures. Especially, competinig order parameter descriptions based on the mass term structure are different.\cite{igor} 
   
{\it Physical Realization :}
One of the well-known examples of the QBT fermion is a bi-layer graphene with Bernal stacking.  
Also, hole-doped GaAs with the inversion symmetry along the perpendicular axis has similar structure. 
In this paper, let us  focus on the bi-layer graphene. 
Its tight binding Hamiltonian is 
\begin{eqnarray}
H_0& =& - t \sum_{<i,j>} (c_{1,i}^{\dagger} c_{1,j} +c_{2,i}^{\dagger} c_{2,j}) + \tilde{t} \sum_{i \in A-B} (c_{1,i}^{\dagger} c_{2,i}  +h.c.) \nonumber \\
\end{eqnarray}
, where $c^{\dagger}_{1(2),i}$ is a creation operator of the first(second) layer at the site(i), 
and the second summation is only for one sublattice where sites on one layer is on top of sites on the other layer, $(A-B)$. 

Among four band dispersions, only two bands become important and the low energy Hamiltonian becomes 
\begin{eqnarray}
H_{G,0} = \psi^{\dagger}\left[ \frac{ (k_x^2 -k_y^2) \sigma^x  +(2 k_x k_y ) \sigma^y \rho^z}{2m}\right] \psi 
\end{eqnarray} 
, where $\sigma, \tau, \rho$ correspond to sub-lattice(or layer), spin, and valley Pauli matrix. 
This Hamiltonian has the helical quadratic dispersion structure as our model Hamiltonian. 
Only considering non-valley mixing states, there are two order parameters with $O(3)$ vector properties. 
\begin{eqnarray}
O_{QSH }=\psi^{\dagger} \sigma^z  \vec{\tau} \rho^z \psi \quad ,  \quad O_{N} =\psi^{\dagger} \sigma^z \vec{\tau} \psi 
\end{eqnarray}
The first one corresponds to the Quantum Spin Hall(QSH) order parameter because it does not break the time reversal symmetry 
even though the spin rotational symmetry is  broken. 
The second one breaks both the spin rotational symmetry and time-reversal symmetry, so it's a Neel order parameter.
To make connnection to our model Hamilnotian manifest, let us rotate the basis as follows. 
\begin{eqnarray}
\psi \rightarrow ( \frac{1+\sigma^x}{2} +\frac{1-\sigma^x}{2} \rho^z ) \psi
\end{eqnarray} 
Then the $\rho$ dependence of the non-interacting Hamiltonian $H_{G,0}$ dropped. 
Of course, the order parameter terms also rotate under the transformation and become
\begin{eqnarray}
O_{QSH} \rightarrow \psi^{\dagger} \sigma^z  \vec{\tau} \psi \quad , \quad O_{N} \rightarrow \psi^{\dagger} \sigma^z \vec{\tau} \rho^z\psi 
\end{eqnarray}
Therefore, the QSH order parameter exactly corresponds to our model Hamiltonain  except doubling of the number of flavors(valley index).

Now we can apply our model Hamiltonian results directly. 
Skyrmion of the QSH order parameter is a boson with charge four considering the valley index. 
On the other hand, if we consider the Neel order parameter, it is easy to see the fermionic charge is zero and the Hopf term becomes trivial because the signs of the coupling constant of different valley fermions are opposite and their contributions are cancelled out.  
It indicates that the Skyrmion of the Neel order parameter is a charge neutral boson. 
Therefore, it is manifest that Skyrmion's induced quantum number depends on order parameters' property, especially how an order parameter couples to Dirac particles with different helicity. 

Let us focus on the QSH phase.
By tuning interactions, one can destroy the QSH order parameter.
Similar to vortex proliferation in $XY$ model in (1+1) dimension, it is possible to eliminate the QSH phase by condensing the Skyrmion due to its bosonic properties. 
This phase cannot be the same as the semimetal phase because the Skyrmion's electric charge is four.
Therefore one can realize the charge $4e$ superconductors by condensing topological defects of the QSH order parameter.
This scenario does not require any Fermi surfaces and the Cooper pair formation, so there is no way to form charge $2e$ order parameters here. 
In other words, we can circumvent the instability problem of the charge $4e$ SC into formation of two charge $2e$ Cooper pairs. 
Therefore, it is one concrete way to achieve charge $4e$ SC through the quantum phase transition from QSH. 

What kind of a quantum phase transition is the transition from QSH to charge $4e$ SC? 
Because two phases break different symmetries, it should be a first order phase transition under the Landau-Ginzburg paradigm. 
But, a second order phase transition is also possible via a deconfined quantum criticality. \cite{senthil, grover}
In quantum mechanics, topological defects of one order parameter can carry the other order parameter's quantum number.\cite{senthil}
Their condensation suppresses the original ordered phase and induces the different symmetry-broken phase simultaneously. 
Thus, a continuous phase transition can naturally appears in this process. 

One candidate critical theory of the phase transition is $O(5)$ model with the Wess-Zumino-Witten term(WZW). \cite{senthil2,pavan}
Introducing a multiplet, $\vec{\phi}=(\vec{n}, {\rm Re \,\psi}_{4e}, {\rm Im \, \psi}_{4e})$, 
the model is 
\begin{eqnarray}
&&S= \int d \tau d^2 x \, \frac{1}{2 g^2}(\partial{\vec{\phi}})^2 + 2\pi i k \Gamma^{WZW}(\vec{\phi}) \\
&&\Gamma^{WZW}(\vec{\phi}) = \nonumber \\
&&\frac{1}{Area(S^{4})} \frac{1}{(4)!}\int_{S^{4}} \epsilon^{ijkl} \epsilon_{abcde} {\phi}^a \partial_{i} {\phi}^b \partial_{j} {\phi}^c \partial_{k} {\phi}^d \partial_{l} {\phi}^e \nonumber 
\end{eqnarray} 
, where $k$ is an integer number level. 

Without the WZW term, the beta function of $g$ is well understood.
There are two fixed points; $g=0$ corresponds to the ordered phase and $g=g_c$ describes a critical fixed point. 
Beyond the critical fixed point, the beta function monotonically increases. 
One can expect the WZW term plays non-trivial effect due to its topological orgin. 
Indeed Witten showed the WZW term in $(1+1)$ dimension induces stable fixed points beyond an unstable fixed point.\cite{witten2}

Here, we follow his method and do the beta function calculation perturbatively. 
We find that 
\begin{eqnarray}
&&\frac{d \tilde{g}^2}{d l} = \beta_0(\tilde{g}^2)- C_0 k^2 \tilde{g}^{10} +O(\tilde{g}^{12},k^4).
\end{eqnarray}
The $\beta_0$ function is one without the WZW term, which is obtained from the literature.\cite{loops2}
\begin{eqnarray}
&&\beta_0(\tilde{g}^2)=- \tilde{g}^2+3 \tilde{g}^4 + 3 \tilde{g}^6 + \frac{21}{4} \tilde{g}^8 + 3 b \, \tilde{g}^{10} \nonumber
\end{eqnarray}
,where $b=\frac{17}{4} + 3 \zeta(3)$ 
and dimensionless coupling constant $\tilde{g}$ 
is introduced to match higher loop calculation.

We note that our calculation is not controlled basically 
because of existence of the original mass scale of the coupling constant. 
In contrast, $O(4)$ NLSM on $1+1$ dimension calculation is controlled by $1/k$ expansion, 
and one can show that there are stable fixed points with non-zero $k$ values, the so-called $SU(2)_k$ WZW fixed point.\cite{witten2}
One physical realization of such fixed points is spin $1/2$ chain.\cite{haldane,affleck} 
Even though our calculation is not controlled, it clearly shows that the topological term induces a non-trivial effect. 
The topological term's contribution is of the opposite sign to the other terms in the beta function. 
Physically, this means that the topological term suppresses the tendency of becoming a disordered state.
 
There are three possible scenarios based on our perturbative calculation. 
First, the topological term contribution is not big enough, so it does not induce any stable fixed point. 
With a small enough value of $k$, the beta function is barely affected by the topological term. 
Phase transition is described by the $O(5)$ NLSM model without the WZW term and  the topological contribution only gives a small correction. 
The second scenario is that the topological term is so big that the beta function becomes always negative. 
An implication of this case is that the system is always in the ordered phase.
Both cases can be described by the $O(5)$ NLSM without the WZW term, so there is no new physics. 
The final case is that the topological term and the higher order fluctuation of the order parameter have similar amounts of effects.
Thus, one new stable fixed point appears beyond the usual critical fixed point.
 
\begin{figure}
\includegraphics[width=3.0 in]{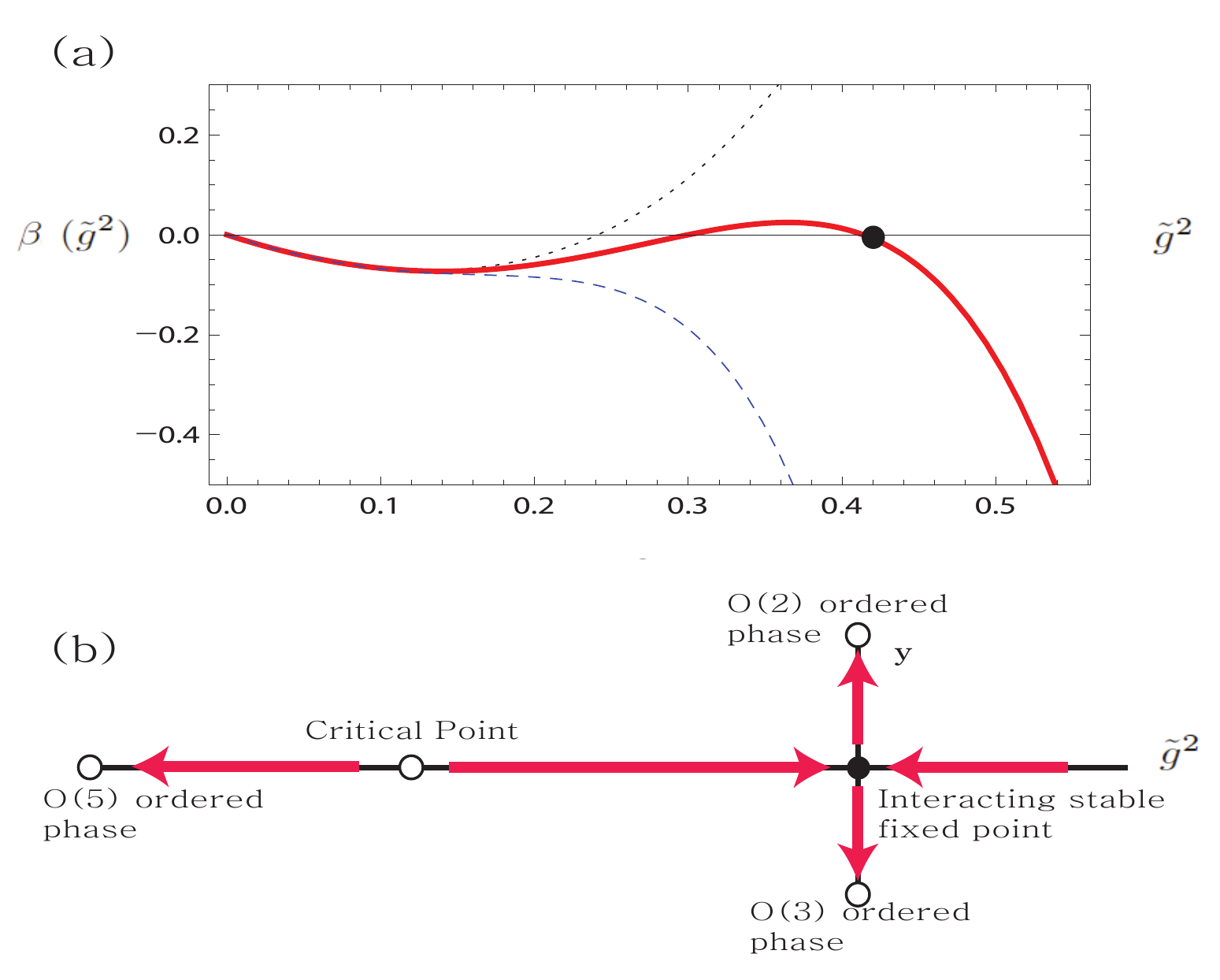}
\caption{(Color Online) (a) Beta function behaviors of $O(5)$ NLSM model with different levels WZW level, $k$. The dotted(black), normal(red), and dashed(blue) lines correspond to $k <k_c, k=k_c, k >>k_c$ cases.
At $k=k_c$, there is one stable interacting fixed point(balck dot). (b) Phase diagram with RG flows for the critical level at $k=k_c$. }
\label{beta}
\end{figure}

This new fixed point is our candidate for the deconfined quantum criticality between QSH and charge $4e$ SC.
By introducing symmetry breaking term such as $ y( \vec{n} \cdot \vec{n} -|\psi_{SC}|^2)$, 
each phase can be accessed from the stable fixed point. 
Numerically, we find $C_0=\frac{ \pi^3}{2}$ and $k_c =2$ with a hard-cutoff regularization scheme.\cite{kc} 
If one believes that the $SU(2)_k$ WZW fixed points in $1+1$ dimension have its extension to ones in the $2+1$ dimension, 
our new fixed point would be a natural candidate. 
Figure \ref{beta} illustrates both the three possible cases in terms of renormalization group flow and the phase diagram at the critical level.

Finally, QBT fermions are different from two copies of Dirac fermions in terms of  fermion mass term structures, which usually explain competing order parameters.
As shown by Herbut\cite{igor} recently, there is a ``hidden'' internal structure(pseudo-spin) of competing order parameters of Dirac particles. 
For mono-layer graphene with Dirac particles, SC order parameter can make competing $O(5)$ vector with either Quantum spin Hall or density wave order parameters.
In other words, there are two possible ways to make competing $O(5)$ order parameters. 
However, one can easily check that QBT fermions do not allow the same two possibilities. Only one choice is allowed. 
For the SC order parameter, only the QSH order parameter is possible to make $O(5)$ vector. 
In that sense, QBT fermions describe a more restricted competing orders' structure, which originates from the difference of Clifford algebra representations of Dirac particles and QBT fermions.\cite{igor} 
We leave this interesting problem for the future work.

{\it Conclusion:} In this paper, we showed how Skyrmions interacting with QBT fermions obtains their quantum numbers.
It is shown that QBT fermions double induced charges compared to Dirac fermions' and make all the Skyrmions bosonic. 
Applying the mechanism to bi-layer graphene, we also showed that QSH Skyrmions in bi-layer graphene are charge $4e$ bosons.
By condensing the Skyrmions, it is possible to achieve charge $4e$ superconductivity.
Such mechanism is far different from the general BCS-like description, and we argue that it might be described by the deconfined quantum criticality.
We propose the $O(5)$ non-linear sigma model with the WZW term at level $k_c$ as a critical theory.

{\it Acknowledgement :} The author thanks to Igor Herbut, Laing Fu, and Cenke Xu for invaluable dicussion.
The author is especially grateful to Cenke Xu for introducing the subject.


\begin{thebibliography}{99}
\bibitem{kc} We define $k_c$ as the smallest value of $k$ to have the stable non-zero fixed point. 
It is difficult to obtain the exact value of $k_c$ which might depend on regularization schemes due to its uncontrolled property.
\end{thebibliography}
\end{document}